\documentclass[twocolumn,showpacs,amsmath,amstex,amssymb,mathfonts]{revtex4}
\usepackage{graphicx,epstopdf,verbatim,color,ulem,amsmath}
\usepackage[braket, qm]{qcircuit}
\DeclareMathOperator{\Tr}{Tr}

\usepackage{color}

\begin{document}
\title{Entanglement spectroscopy on a quantum computer}
\author{Sonika Johri$^{1}$, Damian S. Steiger$^{2}$ and  Matthias Troyer$^{2,3}$}
\affiliation{$^1$ Intel Labs, Intel Corporation, Hillsboro, OR 97124}
\affiliation{$^2$ Theoretische Physik, ETH Zurich, 8093 Zurich, Switzerland}
\affiliation{$^3$ Quantum Architectures and Computation Group, Microsoft Research, Redmond, WA (USA)}

\begin{abstract}
	We present a quantum algorithm to compute the entanglement spectrum of arbitrary quantum states.  The interesting universal part of the entanglement spectrum is typically contained in the largest eigenvalues of the density matrix which can be obtained from the lower Renyi entropies through the Newton-Girard method. Obtaining the $p$ largest eigenvalues ($\lambda_1>\lambda_2\ldots>\lambda_p$) requires a parallel circuit depth of $\mathcal{O}(p(\lambda_1/\lambda_p)^p)$ and $\mathcal{O}(p\log(N))$ qubits where up to $p$ copies of the quantum state defined on a Hilbert space of size $N$ are needed as the input.	
	 We validate this procedure for the entanglement spectrum of the topologically-ordered Laughlin wave function corresponding to the quantum Hall state at filling factor $\nu=1/3$. Our scaling analysis exposes the tradeoffs between time and number of qubits for obtaining the entanglement spectrum in the thermodynamic limit using finite-size digital quantum computers.  We also illustrate the utility of the second Renyi entropy in predicting a topological phase transition and in extracting the localization length in a many-body localized system.
\end{abstract}

\date{\today}
\maketitle
One important application of quantum computers is efficiently simulating many-body quantum systems. While a variety of methods has been advanced for efficiently evolving a quantum state on a quantum computer, extracting useful information from a system of qubits is not always as straightforward.  The quantum computing equivalent of the vast array of diagnostic tools that extract information from classical numerical simulation are still being developed\cite{knill2007}. A recent paper addresses this paucity by developing efficient techniques to estimate expectation values of arbitrary observables and static and dynamic correlation functions using a quantum computer \cite{wecker2015}.

In this paper we address the calculation of quantities which characterize entanglement between different parts of a quantum state using a quantum computer. We assume that the state under investigation may be efficiently prepared by any one of many available techniques for quantum state preparation \cite{grover2002,kaye2004,zalka1998,ward2009,wang2009}. It may also be obtained as the ground state of a Hamiltonian \cite{omalley2015,aspuru-guzik2005,jones2012,veis2010,wang2008,wecker2013,whitfield2011} or as the result of an adiabatic evolution \cite{aharanov2003,babbush2013,farhi2001,perdomo-ortiz2012}. 

To begin,  consider a many-body quantum system composed of two subsystems A and B. Then a wave function $|\psi\rangle$ defined over the Hilbert spaces of A and B can be written as
\begin{align}\label{eq:wfn}
|\psi\rangle=\sum_{ij}c_{ij}|a_i\rangle\otimes|b_j\rangle,
\end{align}
where the states $|a_i\rangle$ and $|b_j\rangle$ form orthonomal bases of A and B, respectively. 
Note that the division into A and B can be in any basis including real space, momentum space, or Fock space.
The reduced density matrix for subsystem A is defined by tracing over the degrees of freedom of B,
\begin{align}
\rho_A=\Tr_B(|\psi\rangle\langle\psi|).
\end{align} 
It contains information about the entanglement between A and B. Using $\rho_A$, we can define the $n$th Renyi entropy, 
\begin{align}
S_{n}=\frac{1}{1-n}\log(R_n)
\end{align}
where
\begin{align}
R_{n}=\Tr(\rho^n_A).
\end{align}
For a generic many-body wave function (that is, not a product state), subsystems A and B will be entangled. For non-zero entanglement, $R_2<1$. $S_2$ has the same universality properties as the von Neumann entropy $S=-\Tr(\rho_A\log(\rho_A))$. They are both non-zero only for entangled subsystems A and B and increase with growing entanglement. These quantities provide valuable information about the underlying physics of the system. For example, whether the entanglement obeys an area law or volume law \cite{eisert2010}, and its evolution with time will determine whether the phase is conducting or insulating \cite{bauer2013,iyer2013,bardarson2012,vosk2013,burrell2007}. It has been used to probe topological order \cite{kitaev2006,wen2006}, quantum critical systems \cite{vidal2003}, and to determine whether classical computers can efficiently simulate particular quantum systems \cite{schuch2008}. 

Li and Haldane \cite{li2008} introduced the concept of the entanglement spectrum which is the energy spectrum of the ``entanglement Hamiltonian" $H_E$ defined through $\rho_A=\exp(-H_E)$. They pointed out that the largest eigenvalues of $\rho_A$ \cite{white1993} contain more universal signatures than the von Neumann entropy or $S_2$ alone. The entanglement spectrum has been used to identify topological order \cite{pollmann2010,fidkowski2010,yao2010} such as the Haldane phase and fractional quantum Hall effect, in systems with broken symmetry \cite{cirac2011,metlitski2011,alba2012,kolley2013,poilblanc2010}, quantum critical systems \cite{calabrese2008}, many-body localization \cite{yang2015,yang2017,serbyn2016}, covalent bonds in molecules \cite{tubman2014}, and irreversibility in quantum systems \cite{chamon2014}. In classical simulations of many-body quantum systems, entanglement entropy and the entanglement spectrum can be extracted from matrix diagonalization, density matrix renormalization group calculations \cite{hu2012}, quantum Monte Carlo simulations \cite{hastings2010,grover2013,chung2014} and other approaches.

Here we present a quantum algorithm to compute the entanglement spectrum via the Renyi entropies on a quantum computer. We calculate $S_n$ by generalizing the swap trick \cite{horodecki2002} which has  recently been used in quantum Monte Carlo calculations, in experiments on ultracold atoms \cite{Islam2015}, and proposed in solid-state spin arrays\cite{banchi2017}.  Next we show how to obtain the low-lying levels of the entanglement spectrum using the Newton-Girard method. We then use the Laughlin wave function which describes the quantum Hall state at filling factor $\nu=1/3$ to validate the procedure, showing that entanglement spectrum levels varying over several orders of magnitude can be extracted given enough accuracy in determining $R_n$. We do a scaling study to analyze the trade-offs between time and number of qubits in obtaining the entanglement spectrum in the {\it thermodynamic} limit using {\it finite-size} digital quantum computers. We also show that the second Renyi entropy itself can be used to predict a topological phase transition, and is also of use in extracting the localization length in a many-body localized system.

We begin with the design of the quantum circuit for $R_2$. 
Let $\psi \in \mathcal{H}$ be a wave function in a Hilbert Space composed of the two subspaces $A$ and $B$, i.e. $\mathcal{H}=A\otimes B $  and $\psi$ as in Eq.~\eqref{eq:wfn}.
We need two copies of the wave function $\psi$ to calculate $R_2$ which is equal to the expectation value of the $\text{Swap}_A$ operator for a system prepared in state $|\psi\rangle|\psi\rangle $, i.e.
\begin{align}
R_2=\langle\psi|\langle\psi|\text{Swap}_A|\psi\rangle|\psi\rangle 
\end{align}
where the operator $\text{Swap}_A$ acts as follows
\begin{align}
\nonumber
\text{Swap}_A |\psi\rangle|\psi\rangle &=\text{Swap}_A\sum_{i,j}c_{ij}|a_i\rangle|b_j\rangle\sum_{i',j'}c_{i'j'}|a_{i'}\rangle|b_{j'}\rangle\\
&=\sum_{i,j}\sum_{i',j'}c_{ij}c_{i'j'}|a_{i'}\rangle|b_j\rangle |a_i\rangle|b_{j'}\rangle \; .
\end{align}
\begin{figure}
\mbox{
\Qcircuit @C=2em @R=1.7em {
\lstick{\ket{0}} & \gate{H} &\ctrl{8} &\ctrl{9} &\ctrl{10} & \gate{H} & \meter\\
&\ustick{\alpha_1}    \qw      &\qswap   &\qw      &\qw      &\qw       & \qw \\
&\ustick{\alpha_2}    \qw      &\qw      &\qswap   &\qw      &\qw       & \qw \\
&\ustick{\alpha_3}    \qw      &\qw      &\qw      &\qswap   &\qw       & \qw\\
&\ustick{\beta_1}    \qw      &\qw      &\qw      &\qw      &\qw       & \qw \\
&\ustick{\beta_2}    \qw      &\qw      &\qw      &\qw      &\qw       & \qw \\
&\ustick{\beta_3}    \qw      &\qw      &\qw      &\qw      &\qw       & \qw \\
&\ustick{\beta_4}    \qw      &\qw      &\qw      &\qw      &\qw       & \qw
\inputgroupv{2}{8}{.8em}{5em}{\ket{\psi}}\\
&\ustick{\alpha_1'}   \qw      &\qswap   &\qw      &\qw      &\qw       & \qw \\
&\ustick{\alpha_2'}   \qw      &\qw      &\qswap   &\qw      &\qw       & \qw \\
&\ustick{\alpha_3'}   \qw      &\qw      &\qw      &\qswap   &\qw       & \qw \\
&\ustick{\beta_1'}   \qw      &\qw      &\qw      &\qw      &\qw       & \qw \\
&\ustick{\beta_2'}   \qw      &\qw      &\qw      &\qw      &\qw       & \qw \\
&\ustick{\beta_3'}   \qw      &\qw      &\qw      &\qw      &\qw       & \qw \\
&\ustick{\beta_4'}   \qw      &\qw      &\qw      &\qw      &\qw       & \qw
\inputgroupv{9}{15}{.8em}{5em}{\ket{\psi}}
}
}
\caption{Quantum circuit to calculate $R_2$ for a quantum state with a Hilbert space that spans 7 qubits. The qubits labelled $\alpha_i$ are in the subsystem A and those labelled $\beta_i$ are in subsystem B. The operator $\text{Swap}_A$ is implemented using three controlled swap gates between qubits $\alpha_i$ and $\alpha_i'$.}
\label{fig:qcircuit_R2}
\end{figure}
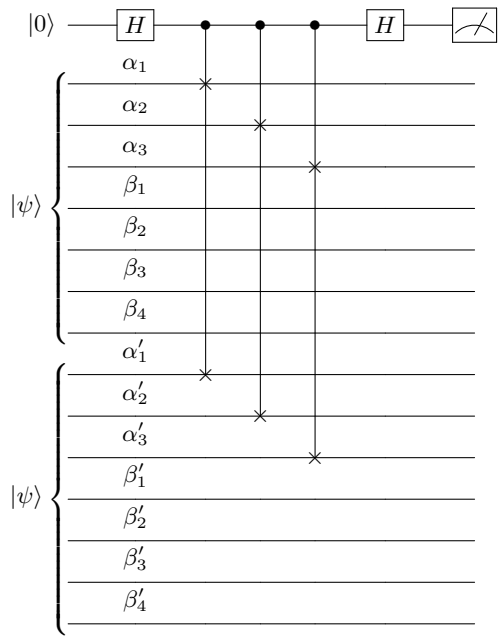
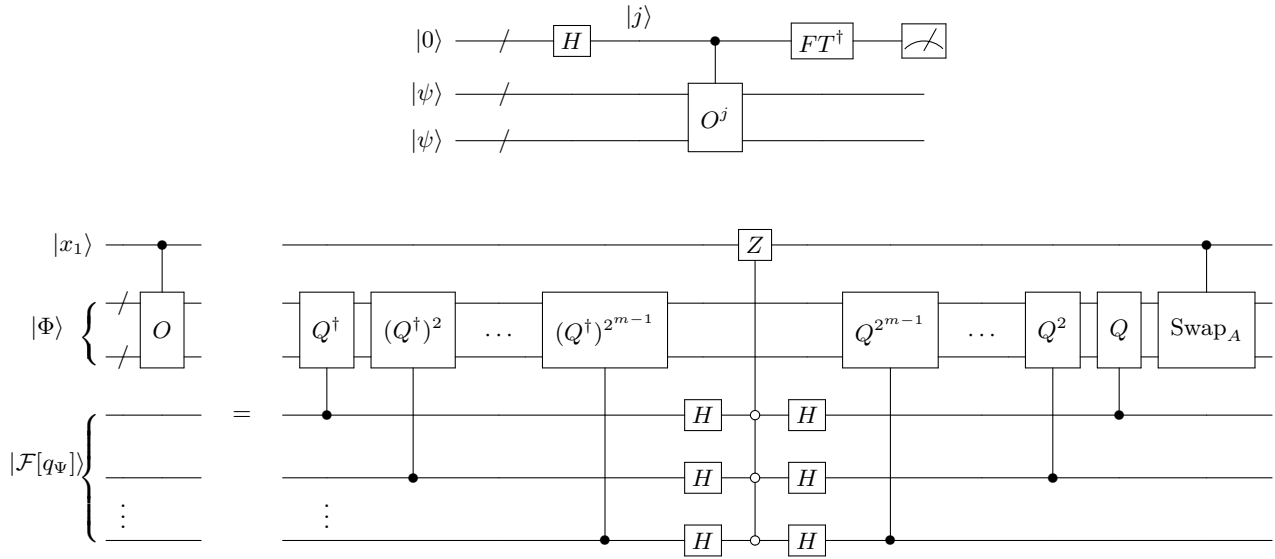
\begin{figure*}
	\centering
\mbox{
	\Qcircuit @C=2em @R=1em {
		&\lstick{|0\rangle}    & {/} \qw & \gate{H} & \ustick{|j\rangle} \qw & \ctrl{1}   & \gate{FT^\dagger} & \meter \\
		&\lstick{|\psi\rangle} & {/} \qw & \qw      & \qw                    & \multigate{1}{O^j} & \qw               & \qw \\
		&\lstick{|\psi\rangle} & {/} \qw & \qw      & \qw                    & \ghost{O^j} & \qw               & \qw
	}
}
\vspace{1cm}

\mbox{
	\Qcircuit @C=.7em @R=1.3em {
		&\lstick{\ket{x_1}} &\qw &\ctrl{1} &\qw & & &\qw &\qw &\qw &\qw &\qw &\gate{Z} &\qw &\qw & \qw & \qw &\qw &\ctrl{1} &\qw\\
		& & {/} \qw &\multigate{1}{O} &\qw & &   & \multigate{1}{Q^{\dagger}} & \multigate{1}{(Q^{\dagger})^2} &\qw & \multigate{1}{(Q^{\dagger})^{2^{m-1}}} & \qw  &\qw & \qw  &\multigate{1}{Q^{2^{m-1}}} &\qw & \multigate{1}{Q^2} & \multigate{1}{Q} & \multigate{1}{\text{Swap}_A} &\qw \\
		& & {/} \qw  &\ghost{O} &\qw & &  & \ghost{Q^{\dagger}} & \ghost{(Q^{\dagger})^2} &\cds{-1}{\cdots} & \ghost{(Q^{\dagger})^{2^{m-1}}}  & \qw & \qw &\qw &\ghost{Q^{2^{m-1}}} &\cds{-1}{\cdots} &\ghost{Q^2} &\ghost{Q} &\ghost{\text{Swap}_A} &\qw 
		\inputgroupv{2}{3}{0em}{1em}{\ket{\Phi}}\\
		& &\qw & \qw &\qw &\push{\rule{.3em}{0em}=\rule{.3em}{0em}} &   & \ctrl{-1} & \qw  & \qw  & \qw & \gate{H} &\ctrlo{-3} & \gate{H} & \qw & \qw &\qw &\ctrl{-1} &\qw & \qw  \\
		& &\qw & \qw &\qw & & & \qw & \ctrl{-2}  & \qw  & \qw   & \gate{H} &\ctrlo{-1} & \gate{H} &\qw &\qw &\ctrl{-2} &\qw &\qw & \qw  \\
		& &\ustick{\vdots} \qw & \qw &\qw & & &\ustick{\vdots} \qw & \qw  & \qw  & \ctrl{-3} & \gate{H}  &\ctrlo{-1} & \gate{H} &\ctrl{-3} &\qw &\qw & \qw &\qw &\qw 
		\inputgroupv{4}{6}{0em}{2em}{\ket{\mathcal{F}[q_{\Psi}]}}
	}
}

\caption{Quantum amplitude estimation to calculate $R_2$. Top: Quantum phase estimation on the operator $O=\text{Swap}_AV$ will give $R_2$. $FT^\dagger$ refers to the inverse Fourier Transform operator. Bottom: Quantum circuit showing controlled implementation of the operator $O$. The operator $Q$ has the desired wavefunction as an eigenstate with the corresponding eigenvalue $q_{\Psi}$ known a priori. The Fourier Transform of this value $\mathcal{F}[q_{\Psi}]$ is stored in the ancilla qubits that remain unchanged at the end of the computation. $Z$ refers to the Pauli-Z gate.}
\label{fig:R2_qae_circuit}
\end{figure*}
The quantum circuit for measuring $R_2$ uses two copies of the state $|\psi\rangle$ prepared in a basis that encodes the two subspaces A and B using distinct sets of qubits, see Fig.~\ref{fig:qcircuit_R2}. The eigenvalues of the swap operator are $\pm 1$ and we need only a single ancilla qubit for a straightforward measurement of its expectation value. Here the ancilla qubit is put into a superposition by the Hadamard gate $H=\frac{1}{\sqrt{2}}\bigl[ \begin{smallmatrix}1 & 1\\ 1 & -1\end{smallmatrix}\bigr]$. Repeated measurements in this manner will result in convergence to the mean with an  accuracy of $\epsilon\sim 1/\sqrt{N_{\text{meas}}}$, where $N_{\text{meas}}$ is the number of measurements. The technique of quantum amplitude estimation (QAE) \cite{brassard} can be used to improve the time scaling. It requires an operator $Q$ which has $\ket{\Psi}$ as an eigenstate with the corresponding eigenvalue $q_{\Psi}$ known a priori. An ancilla register stores the Fourier Transform of this value $\mathcal{F}[q_{\Psi}]$. The idea is to implement an operator $V=1-2\ket{\Psi}\bra{\Psi}$ using a projector onto $\ket{\Psi}=\ket{\psi}\ket{\psi}$. The eigenvalues of $O=\text{Swap}_AV$ are $-\exp(\pm 2i\theta)$ where $\cos^2(\theta)=(\bra{\Psi}\text{Swap}_A\ket{\Psi}+1)/2$. Quantum phase estimation for the operator $O$ will produce the value $\theta$ to required accuracy in one run of the circuit (Fig.~\ref{fig:R2_qae_circuit}). The number of qubits required to store the value of $\theta$ will be of $\mathcal{O}(\log(1/\epsilon))$. The controlled application of the operator $O$ effects the following transformation:
\begin{align}
&\ket{x}\ket{\Phi}\ket{\mathcal{F}[q_{\Psi}]}\nonumber\\
\rightarrow & \ket{x}\bigg(c_{\Psi}\ket{\Psi}\ket{0}+\sum_{i\neq 0} c_i\ket{\phi_i}\ket{i}\bigg)\nonumber\\
\rightarrow & \ket{x}\bigg((-1)^x c_{\Psi}\ket{\Psi}\ket{0}+\sum_{i\neq 0} c_i\ket{\phi_i}
\ket{i}\bigg)\nonumber\\
\rightarrow&\ket{x} (\ket{\Phi}-2xc_{\Psi}\ket{\Psi})\ket{\mathcal{F}[q_{\Psi}]}\nonumber\\
\rightarrow& \ket{x}(\text{Swap}_A(\ket{\Phi}-2xc_{\Psi}\ket{\Psi}))\ket{\mathcal{F}[q_{\Psi}]}
\end{align}
Here $x$ is the computational basis state of the control qubit, either 0 or 1. $O$ acts on $\ket{\Phi}$ which has overlap $c_{\Psi}$ with $\ket{\Psi}$. In the first step, the inverse of quantum phase estimation with the operator $Q$ will send $\mathcal{F}[q_{\Psi}]$ to $0$ if the input to $O$ is $\Psi$ and to a non-zero number otherwise. The run-time for quantum amplitude estimation is $\mathcal{O}(T_Q\epsilon_Q^{-1}\epsilon^{-1})$, where $T_Q$ is the time to implement a control-$Q$ gate, and $\epsilon_Q$ is the difference between $q_{\Psi}$ and the closest other eigenvalue to it of $Q$. 
 
We can generalize the above method to calculate $R_n$ for $n>2$ by using the permutation operator on the tensor product of $n$ copies of the wave function:
\begin{align}
\text{Perm}_A|\psi\rangle^{\otimes n}=\text{Swap}_A^{(n-1)\leftrightarrow n}
\dots\text{Swap}_A^{3\leftrightarrow 2}\text{Swap}_A^{2\leftrightarrow 1}|\psi\rangle^{\otimes n}
\end{align}
Then,
\begin{align}
\begin{split}
R_n & = \Tr (\rho_A^n) = \Tr \left( \left( \sum_{ijk}c_{ij}c^*_{kj}|a_i\rangle \langle a_j| \right)^n \right)\\ 
& =\sum_{\text{all indices}} \gamma_{i_1 i_2}\gamma_{i_2 i_3}\dots\gamma_{i_n i_1} 
=\langle\psi|^{\otimes n}\text{Perm}_A|\psi\rangle^{\otimes n}
\end{split}
\end{align}
where $\gamma_{ik}=\sum_{j}c_{ij}c^*_{kj}$ and $c_{ij}$ are used in the definition of the wave function $|\psi\rangle$ from Eq. \ref{eq:wfn}.

To measure $R_n$ we prepare $n$ copies of the wave function and calculate the expectation value of the permutation operator. 
Time efficiency can be improved by using QAE for the permutation operator as well similarly to how it is used for the swap operator.

The number of qubits needed for calculating $R_n$ scales as $\mathcal{O}(n\log(N))$, where $N$ is the size of the Hilbert space on which $|\psi\rangle$ is defined. The number of gates for a single measurement will scale as $\mathcal{O}(n\log(N_A))$, with $N_A$ the size of the Hilbert space of sub-system A.

The calculation of the entanglement spectrum would at first glance seem to require the computation of all the elements of the density matrix $\rho_A$. From the series expansion,
\begin{align}
R_n=\sum_i \lambda_i^n,
\end{align}
where $\lambda_i$ are the eigenvalues of $\rho_A$, we see that calculating all the $R_n$ is equivalent to finding all the eigenvalues. However, for many Hamiltonians of interest, the eigenvalues of the entanglement Hamiltonian will differ by several orders of magnitude and only largest few eigenvalues are interesting. For example, to distinguish between the possible conformal field theories associated with some fractional quantum Hall wave functions, it is sufficient to have access to between one and ten of the largest eigenvalues in each momentum sector even for system sizes approaching the thermodynamic limit \cite{li2008}.
These large eigenvalues can be estimates from just a few $R_n$, for small $n$. 

We use the Newton-Girard method \cite{song2012} which relates the coefficients of the characteristic polynomial of a square matrix ($\rho_A$) of size $N_A$ to the power sums ($R_n$) of its roots. Briefly,
\begin{align}
(x-\lambda_1)(x-\lambda_2)...(x-\lambda_{N_A})=\sum_{k=0}^{N_A} (-1)^{n+k} e_{n-k} x^{k}
\end{align}
Then,
\begin{align}
\nonumber e_0 &= 1,\\
\nonumber e_1 &= R_1,\\
\nonumber e_2 &= \frac{1}{2}(e_1 R_1 - R_2),\\
\nonumber e_3 &= \frac{1}{3}(e_2 R_1 - e_1 R_2 + R_3),\\
\nonumber e_4 &= \frac{1}{4}(e_3 R_1 - e_2 R_2 + e_1 R_3 - R_4),\\
& {} \  \  \vdots
\end{align}
We can truncate the polynomial to order $n_{\max}$ where $n_{\max}$ is the highest order Renyi entropy we are able to calculate. The roots of the truncated polynomial will give an approximation to the $n_{\max}$ largest eigenvalues of $\rho_A$.

We now turn to a test case to validate the procedure discussed above. For this, we use a wave function which represents a fractional quantum Hall effect (FQHE) state. FQHE occurs in two-dimensional electron gases (such as in GaAs-AlGaAs heterojunctions) in the presence of a strong transverse magnetic field at low temperatures \cite{tsui1982,stormer1999}. FQHE states exhibit  plateaus in the Hall resistance at certain rational fractional values of the filling factor $\nu$ (the ratio of electrons to magnetic flux quanta), which are accompanied by the vanishing of the longitudinal resistance. The topological order in FQHE states can often be identified from the low-lying levels of the entanglement spectrum and is related to the spectrum of the associated conformal field theory. Thus it can be used as a ``fingerprint" for identifying topological order in wave functions.   

To simulate the FQHE state at $\nu=1/3$, we confine electrons to the $x$-$y$ plane in a magnetic field $B\hat{z}$. We work in the Landau gauge with vector potential $\vec{A}=Bx\hat{y}$. This makes the momentum along the $y$ direction, $k_y$, a good quantum number. The single particle wave functions or ``orbitals'' in the lowest Landau level have the form
\begin{align}\label{eq:onebody}
\phi_{k_y}(\vec{r})=\frac{1}{\pi^{1/4}\sqrt{L}}\exp\bigg(ik_yy-1/2(x/l_B-k_y l_B)^2\bigg),
\end{align}
where $l_B=\sqrt{\hbar/eB}$ is the magnetic length, which we set to 1 below. We use periodic boundary conditions along the $y$-axis at $y=0$ and $y=L$ implying that the electrons live on the surface of a cylinder as shown in Fig. 3. The allowed values of $k_y$ are $2\pi m/L$, where $m$ is an integer. We use a finite number of orbitals $N_{\text{orb}}$ which is set by the filling factor. In the cylindrical geometry, at $\nu=1/3$, a unique ground state occurs when $N_{\text{orb}}= 3N_e-2$, where $N_e$ is the number of electrons.  The one-body wave functions are thus a product of a Gaussian function along the $x$-axis centered at $x= 2\pi m/L$ and a periodic function along the $y$-axis as shown in Fig. 3.

Instead of the true Coulomb interaction, we first use a short-range interaction
\begin{align}
H_L=\nabla^2\delta(\vec{r}),
\end{align}
in our example, where $\delta(\vec{r})$ is a two-dimensional delta function on the surface of the cylinder. The ground state of $H_L$ is the Laughlin wave function which has greater than $99\%$ overlap with, and captures the topological properties of the ground state of the Coulomb interaction, but is less susceptible to finite-size effects. Analytically it can be  written as
\begin{align}
\Psi_L=\sum_{i<j}(z_i-z_j)^3\exp\bigg(-\sum_k|z_k|^2/4\bigg)
\end{align}
We work in the second-quantized basis in which the many-body wave function can be written in terms of occupations of the one-body orbitals in Eq. \ref{eq:onebody}. The total angular momentum $K=(2\pi/L)\sum_m (n_m-(N_{\text{orb}}-1)/2) m $ remains a good quantum number under the interaction $H_L$.

To calculate the entanglement spectrum, we use the orbital basis to partition the system into regions A and B. The electron number in either region ($N_{eA}$ and $N_{eB}$) and the momentum ($K_{A}$ and $K_B$) are good quantum numbers with the constraint $N_e=N_{eA}+N_{eB}$ and $K=K_A+K_B$. Thus the entanglement spectrum corresponding to $\rho_A$ separates into sectors labeled by $N_{eA}$ and $K_A$. 

\begin{figure}
	\includegraphics[width=\columnwidth]{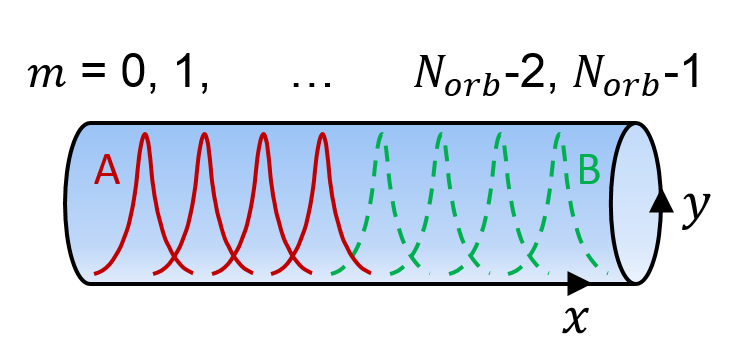}
	\caption{Schematic of cylindrical surface showing $N_{\text{orb}}$ one-body Landau gauge wave functions which are Gaussian along the length of the cylinder ($x$ axis) and periodic along the circular direction. The cut in orbital space preserves momentum along the y-axis of the cylinder.}
\end{figure}
We calculate $R_n$ for $n\le5$ for the ground state of $H_L$ (obtained by exact diagonalization) for upto $N_e=11$ electrons at $\nu=1/3$ and test the feasibility of obtaining the low-lying values of the entanglement spectrum from the truncated characteristic polynomial of the matrix $\rho_A$. This may not be straightforward because the higher $R_n$ will be dominated by the largest eigenvalue since $R_n=\lambda_1^n(1+(\lambda_2/\lambda_1)^n+...)$. Therefore, one may only access the $i$th-largest eigenvalue if $(\lambda_i/\lambda_{\max})^i>\epsilon$, where $\epsilon$ is the accuracy of $R_n$. The blue crosses in Fig. 4 show the entanglement spectrum for $N_e=10$ electrons in $N_{\text{orb}}=28$ orbitals with the cut resulting in $N_{eA}=5$ with equal number of orbitals in A and B. It is clear that the eigenvalues in each momentum sector vary over several orders of magnitude. Rather than the exact values, the number of non-zero eigenvalues in each momentum sector is important here. In order to obtain the maximum number of eigenvalues with maximum accuracy from the Renyi entropies, we find that the following iterative strategy works well. We truncate the characteristic polynomial to order $p$, then calculate $p$ roots using Matlab's root-finding function. If $(\lambda_{\min}/\lambda_{\max})^p<10^{-15}$, we terminate the procedure because we cannot hope to obtain the next smallest eigenvalue accurately. Otherwise, we increase $p$ by 1 and repeat. The results from the algorithm (red circles) faithfully reproduce the results from exact diagonalization (blue crosses) for the lower part of the spectrum. The higher part of the spectrum remains inaccessible because of the limited accuracy ($\sim 10^{-15}$) of the double data type. Using less precision would mean that we obtain fewer levels in each momentum sector. We also point out that relatively low-lying levels such as the 5th level at $K_A=4$ for which even $(\lambda_4/\lambda_1)^4 \sim 10^{-20}$ could be missed by our technique. To remedy this, we need to go to larger system sizes as discussed below.

\begin{figure}
	\includegraphics[width=\columnwidth]{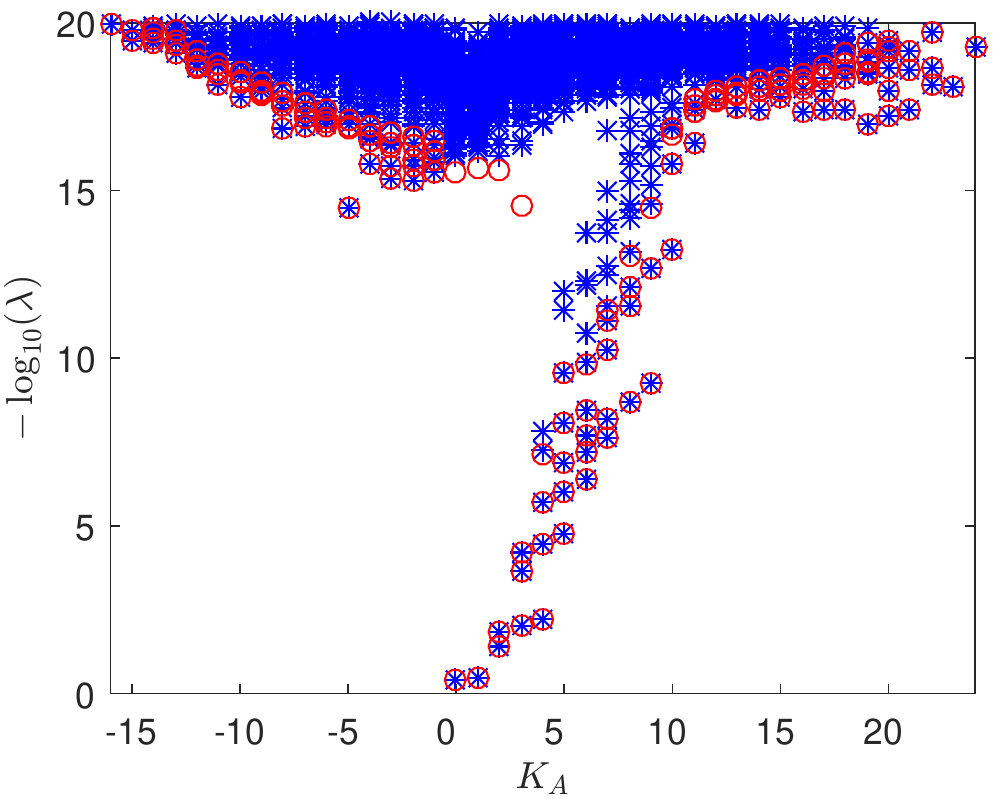}
	\caption{The entanglement spectrum for the ground state of a system of $N_e=10$ electrons in $N_{\text{orb}}=28$ orbitals on the surface of a cylinder. The system is cut in the center with equal number of electrons on each side. Blue crosses correspond to results from exact diagonalization of $\rho_A$ while red circles correspond to results from the algorithm.}
	\label{fig:es}
\end{figure}
While the above procedure provides proof-of-principle that eigenvalues varying over several orders of magnitude can be extracted from $R_n$, the high accuracies required will take an impractical amount of time. However, the way around this is to realize that, for a given momentum sector, as the system size increases, the eigenvalues of the density matrix will become closer together in magnitude and in the thermodynamic limit, they will be degenerate \cite{thomale2010}. Fig. 5 shows the spread $\Delta_3=\log_{10}(\lambda_{\max}/\lambda_{\min})$ in the momentum sector $K_A=3$ which has 3 non-zero eigenvalues as a function of inverse system size. A fit to the data of the form $a(1/N_e)^c+b$ shows that in the thermodynamic limit the splitting goes to zero. The inset shows the corresponding implications for the number of qubits required to represent the wavefunction and the accuracy required by the Renyi entropy technique to get the right number of non-zero eigenvalues. In the thermodynamic limit, when the entanglement spectrum levels are degenerate, the values of $R_n$ will need only to be determined to $\mathcal{O}(1)$ accuracy. For an arbitrary momentum sector, the spread $\Delta=\log(\lambda_{\max}/\lambda_{\min})\sim N_e^{-c}$, where $c>0$. Then the accuracy required is $\epsilon\sim(\lambda_{\min}/\lambda_{\max})^p$ to determine $p$ non-trivial eigenvalues. Therefore, $\epsilon\sim \exp(-N_e^{-c}p)$, and the time required is $\mathcal{O}(p\exp(N_e^{-c}p))$. The space requirements scale as $\sim p N_e$ since the size of the Hilbert space $N\sim 2^{N_e}$ for constant filling factor $\nu$ at large sizes. 

This technique should be compared with the method in Ref. \cite{pichler2016} where the time and space requirements both scale with the accuracy as $1/\epsilon^2$. For finite size systems, with several orders of magnitude splitting between eigenvalues, our technique provides a clear advantage in terms of the number of qubits required. With quantum amplitude estimation, our time scaling is also better.

\begin{figure}
	\includegraphics[width=\columnwidth]{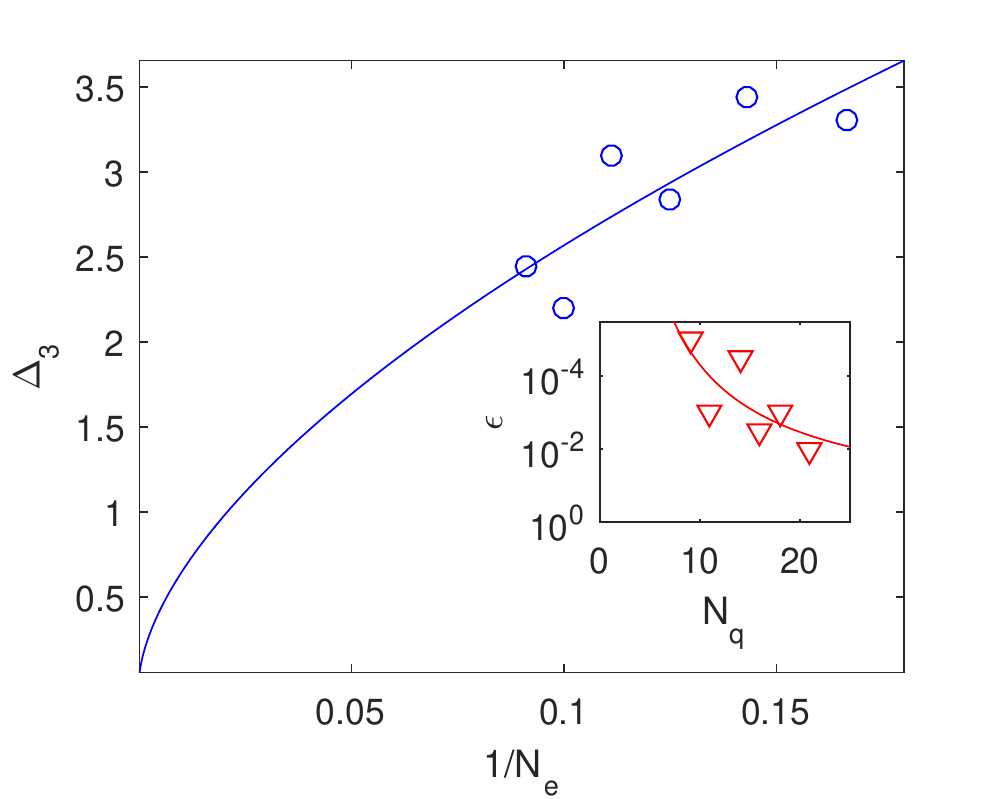}
	\caption{Spread of the entanglement spectrum of the Laughlin wavefunction at $K_A=3$ as a function of system size. The solid line is the least-squares fit of the form $a(1/N_e)^c+b$, with values $a=10.22, b=0, c=0.6$. The inset shows the accuracy required to get the correct number of non-zero eigenvalues at $K_A=3$ versus the number of qubits required to represent the wavefunction. The least-squares fit $\log(\epsilon)=a'N_q^{-c'}$ gives $c'=0.8$ which is close to the value of $c$ obtained from the fit in the main figure as predicted by the arguments in the text.		
	}
	\label{fig:es}
\end{figure}
We next show how the second Renyi entropy $S_2$ can by itself be used to capture a topological phase transition even when it is known only to accuracy of $\mathcal{O}(1)$. Fig. 6 shows the value of $S_2$ diverging at a phase transition between a topologically ordered Laughlin phase and a topologically trivial phase. Here, the Hamiltonian used is $H_c+fH_L$ where $H_c$ is the Coulomb interaction for electrons on the surface of the cylinder described above and $H_L$ is the Laughlin interaction as before. As larger amount of $H_L$ is subtracted from $H_c$, the short-range repulsive component of the interaction disappears leading to the destruction of topological order at a critical value $f_c=0.61$.

\begin{figure}
	\includegraphics[width=\columnwidth]{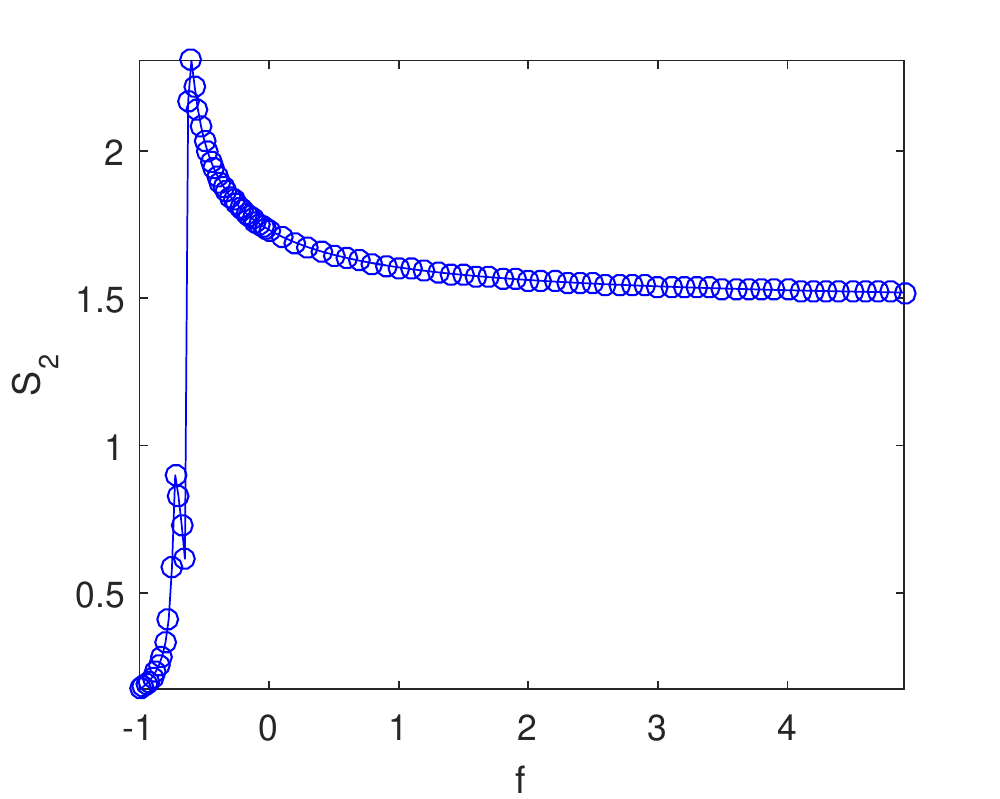}
	\caption{The second Renyi entropy $S_2$ diverges at the phase transition between the Laughlin state and a topologically trivial phase for $N_e=8$ electrons in $N_{\text{orb}}=22$ electrons, cut in the center with equal electrons on either side at $K_A=0$.}
	\label{fig:s2}
\end{figure}
Next, we show that the Renyi entropy can also be used to compute the localization length of a many-body localized (MBL) system. For an eigenstate of a sufficiently large MBL system, the entanglement spectrum decays as a power-law, namely $\lambda_k=ck^{-\gamma}$. $\gamma=4\kappa/\ln(2)$, where $\kappa$ is the many-body localization length, and $c$ is a constant \cite{serbyn2016}. Since we also know that $\sum_k \lambda_k=1$, we can express the second Renyi entropy as a function of $\gamma$:
\begin{align}\label{eq:R2_and_gamma}
R_2=\frac{\sum_k k^{-2\gamma}}{\bigg(\sum_k k^{-\gamma}\bigg)^2}
\end{align}
Thus $R_2$ is a monotonic function of $\gamma$ and a measurement of $R_2$ provides a direct measurement of the many-body localization length without measuring all the components of the wavefunction.

\begin{figure}
	\includegraphics[width=\columnwidth]{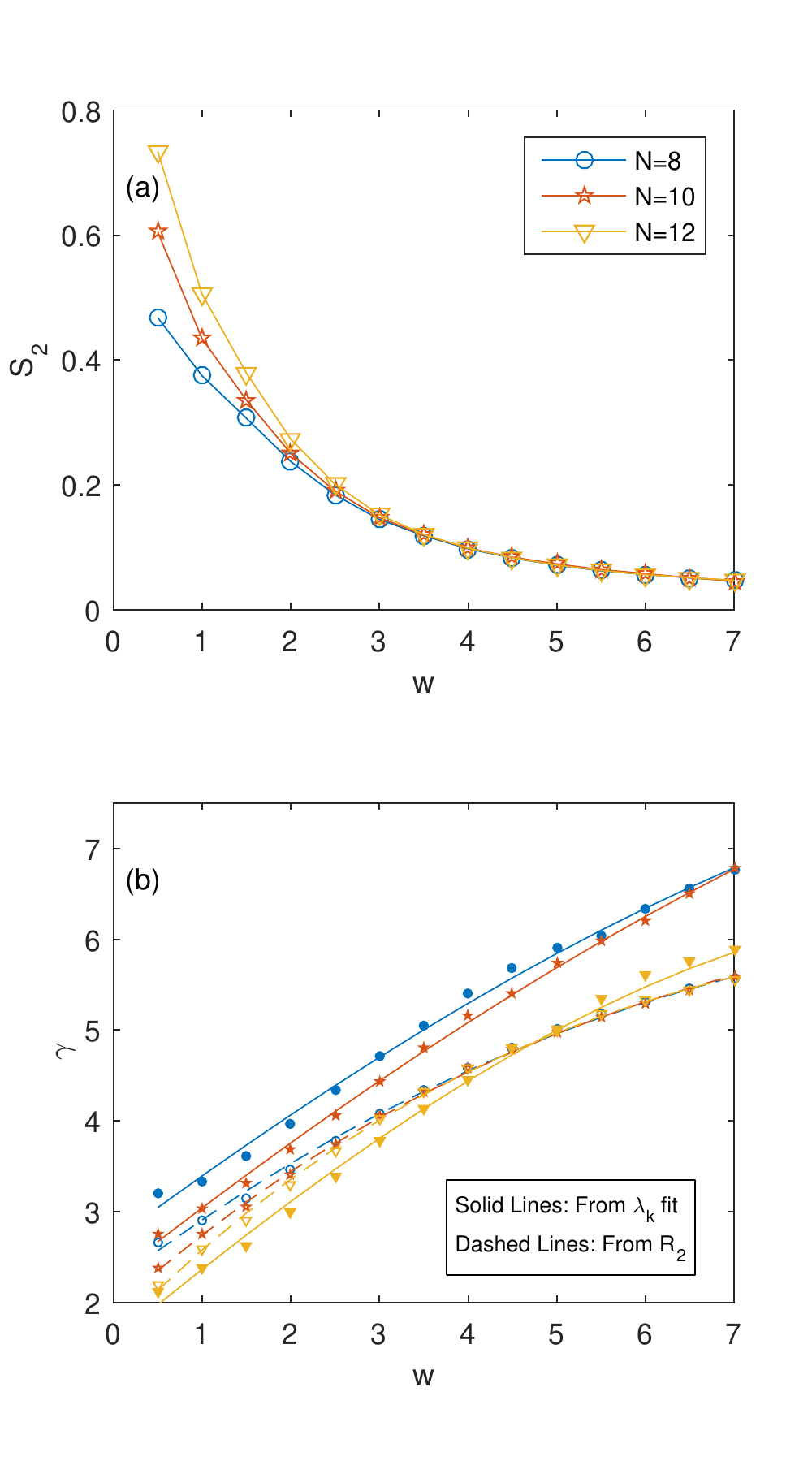}
	\caption{(a) The second Renyi entropy $S_2$ as a function of the disorder parameter $w$ for different system sizes obeying the Hamiltonian in Eq. \ref{eq:MBL}. (b) The parameter $\gamma$ which is a measure of the many-body localization length as extracted from a fit to the first few levels of the entanglement spectrum, and the same as extracted from $R_2$ using Eq. \ref{eq:R2_and_gamma}.}
	\label{fig:MBL}
\end{figure}

We verify this for a standard model of MBL - the antiferromagnetic Heisenberg chain consisting of $N$ spin-$1/2$ sites with random $z$ fields:
\begin{align}\label{eq:MBL}
H_{\text{MBL}}=\sum_{i=1}^{N-1}J\vec{\sigma}_i.\vec{\sigma}_{i+1}+\sum_{i=1}^{N}h_i\sigma_i^z
\end{align}
We set $J=1$ and use a uniform distribution between $-w$ and $w$ for the random fields $h_i$. This model is known to have a many-body localization transition at $w\approx3.5$. We use exact diagonalization to obtain the eigenstates over the entire spectrum for various values of $w$ and disorder realizations with up to 12 sites. Using the eigenstates in the middle third of the spectrum, we disorder-average over 1000 samples to obtain both the entanglement spectrum and $S_2$ with the system being cut in the center.  Fig. \ref{fig:MBL} (a) shows that the value of $S_2$ increases with decreasing disorder around the phase transition. To the right of the critical point, in the localized phase, the Renyi entropy is independent of $N$ indicating area-law entanglement, whereas to the the left, in the thermalized phase, it increases with the system size, indicating volume-law entanglement. We use the formula in Eq. \ref{eq:R2_and_gamma} to obtain $\gamma$. We also obtain $\gamma$ from fitting the first $2^{N/2-1}-1$ entanglement spectrum levels (which are the ones expected to follow power-law behavior according to the arguments in \cite{serbyn2016}) to a straight line. Fig. \ref{fig:MBL} (b) shows the results for $\gamma$ as a function of disorder strength from both these techniques. Both values follow the same trend with greater convergence as the system size increases. Thus we verify the usability of the formula in Eq. \ref{eq:R2_and_gamma} and show that the second Renyi entropy alone is enough to give a good approximation to the many-body localization length. On a quantum computer, $2N+1$ qubits can perform this computation with accuracy required being $\mathcal{O}(1)$.

Thus, in this paper, we have shown how quantum computers can be used to extract the Renyi entropies and the entanglement spectrum, quantities that are relevant to several areas of quantum physics. The entanglement spectrum is entirely a property of the wave function and can thus be used to differentiate between several candidate wave functions even when the Hamiltonian is not available. We have validated the algorithm for the topologically ordered Laughlin wave function and shown that our technique can be used to extract entanglement spectrum levels that are separated by several orders of magnitude. Further, we have analyzed the tradeoffs between time and number of qubits for obtaining the thermodynamic entanglement spectrum on a finite-size digital quantum computer. We have also shown the utility of the second Renyi entropy in studying phenomena as diverse as topological phase transitions and many-body localization. If real quantum computers are to be used on a regular basis, such techniques will be important for their application to studying many-body problems in condensed matter physics. As the next step, we aim to determine quantum gate counts for determining the entanglement spectrum of correlated wavefunctions that describe such systems.

We acknowledge discussions with Zlatko Papic and Alexios Michailidis. DSS and MT have been supported by the Swiss National Science Foundation through the National Competence Center in Research QSIT.

\end{document}